\documentclass[12pt]{article}
\newcommand{\bq}{\begin{equation}}
\newcommand{\eq}{\end{equation}}
\newcommand{\ba}{\begin{eqnarray}}
\newcommand{\ea}{\end{eqnarray}}

\newcommand{\nl }{ \nonumber  }

\newcommand{\p}{\partial}

\newcommand{\h}{\hspace{.5cm}}
\newcommand{\s}{\sigma}

\begin{document}
\begin{flushright}
{\bf hep-th/0103154}
\end{flushright}
\vspace*{1.cm} {\bf\begin{center} EXACT STRING SOLUTIONS IN NONTRIVIAL
BACKGROUNDS \vspace*{0.5cm}\\ P. Bozhilov \footnote {E-mail:
p.bozhilov@shu-bg.net},
\\ \it Department of Theoretical and Applied Physics, \\
Shoumen University "Konstantin Preslavsky", 9712 Shoumen, Bulgaria
\end{center}}
\vspace*{0.5cm}
We show how the classical string dynamics in $D$-dimensional gravity
background can be reduced to the dynamics of a massless particle
constrained on a certain surface whenever there exists at least one
Killing vector for the background metric. We obtain a number of sufficient
conditions, which ensure the existence of exact solutions to the equations
of motion and constraints. These results are extended to include the
Kalb-Ramond background. The $D1$-brane dynamics is also analyzed and exact
solutions are found. Finally, we illustrate our considerations with
several examples in different dimensions. All this also applies to the
tensionless strings.
\\

PACS number(s): 04.25.-g, 11.10.Lm, 11.25.-w, 11.27.+d


\vspace*{.5cm}

\section{Introduction}
\hspace{1cm}The string equations of motion and constraints in curved
space-time are highly nonlinear and, {\it in general}, non exactly
solvable. Different methods have been applied to solve them approximately
\cite{dVS87}, \cite{GSV91}, \cite{dVN92}-\cite{CLS96} or, if possible,
exactly \cite{MR92}-\cite{MO00}. On the other hand, quite general exact
solutions can be found by using an appropriate ansatz, which exploits the
symmetries of the underlying curved space-time \cite{dVS94},
\cite{dVE93}-\cite{dVE95}. In most cases, such an ansatz effectively
decouples the dependence on the spatial world-sheet coordinate $\sigma$
\cite{dVE93}-\cite{LS00} or the dependence on the temporal world-sheet
coordinate $\tau$ \cite{LS96,LS98}, \cite{LS951}-\cite{SM96}. Then the
string equations of motion and constraints reduce to nonlinear coupled
{\it ordinary} differential equations, which are considerably simpler to
handle than the initial ones.

In this article, we obtain some exact solutions of the classical equations
of motion and constraints for both tensile and null strings in a
$D$-dimensional curved background. This is done by using an ansatz, which
reduces the initial dynamical system to the one depending on only one
affine parameter. This is possible whenever there exists at least one
Killing vector for the background metric. Then we search for sufficient
conditions, which ensure the existence of exact solutions to the equations
of motion and constraints without fixing particular metric. These results
are extended to include the Kalb-Ramond two-form gauge field background.
The $D$-string dynamics is also analyzed and exact solutions are found.
After that, we give several explicit examples in four, five and ten
dimensions. Finally, we conclude with a discussion of the derived results.

\section{Reduction of the dynamics}
\hspace{1cm}To begin with, we write down the bosonic string action in $D$-
dimensional curved space-time $\mathcal{M}_D$ with metric tensor $g_{MN}$
\ba\label{as} S&=&\int d^{2}\xi {\cal L},\h {\cal L}=
-\frac{T}{2}\sqrt{-\gamma}\gamma^{mn} \p_m X^M\p_n X^N
g_{MN}\left(X\right),
\\ \nl
\p_m &=&\p/\p\xi^m,\h \xi^m=(\xi^0,\xi^1)=(\tau,\s),\h m,n=0,1,\h
M,N=0,1,...,D-1, \ea where, as usual, $T$ is the string tension and
$\gamma$ is the determinant of the auxiliary metric $\gamma_{mn}$.

Here we would like to consider tensile and null (tensionless) strings on
equal footing, so we have to rewrite the action (\ref{as}) in a form in
which the limit $T\to 0$ can be taken. To this end, we set \ba\label{par}
\gamma^{mn}= \left(\begin{array}{cc}-1&\lambda^1\\
\lambda^1&-\left(\lambda^1\right)^2 + \left(2\lambda^0
T\right)^2\end{array}\right)\ea and obtain \ba\nl {\cal L}=
\frac{1}{4\lambda^0}g_{MN}\left(X\right) \left(\p_0-\lambda^1\p_1\right)
X^M\left(\p_0-\lambda^1\p_1\right) X^N - \lambda^0 T^2
g_{MN}\left(X\right)\p_1X^M\p_1X^N.\ea The equations of motion and
constraints following from this Lagrangian density are \ba\nl \p_0\left[
\frac{1}{2\lambda^0}\left(\p_0-\lambda^1\p_1\right) X^K\right] -
\p_1\left[ \frac{\lambda^1}{2\lambda^0}
\left(\p_0-\lambda^1\p_1\right)X^K\right]\\ \nl + \frac{1}{2\lambda^0}
\Gamma^{K}_{MN}\left(\p_0-\lambda^1\p_1\right) X^M
\left(\p_0-\lambda^1\p_1\right) X^N = \\ \nl 2\lambda^0 T^2 \left[
\p_1^2X^K + \Gamma^{K}_{MN}\p_1X^M\p_1X^N +
\left(\p_1\ln\lambda^0\right)\p_1X^K\right],\\
\label{c1}g_{MN}\left(X\right) \left(\p_0-\lambda^1\p_1\right) X^M
\left(\p_0-\lambda^1\p_1\right) X^N + \left(2\lambda^0 T\right)^2
g_{MN}\left(X\right)\p_1X^M\p_1X^N = 0,\\
\label{c2}g_{MN}\left(X\right)\left(\p_0-\lambda^1\p_1\right)
X^M\p_1X^N=0, \ea where \ba\nl\Gamma^{K}_{MN} =
\frac{1}{2}g^{KL}\left(\p_Mg_{NL} + \p_Ng_{ML} - \p_Lg_{MN}\right)\ea is
the connection compatible with the metric $g_{MN}\left(X\right)$. We will
work in the gauge $\lambda^m = constants$ in which the Euler-Lagrange
equations take the form \ba\label{ele}\left(\p_0-\lambda^1\p_1\right)
\left(\p_0-\lambda^1\p_1\right)X^K + \Gamma^{K}_{MN}
\left(\p_0-\lambda^1\p_1\right) X^M \left(\p_0-\lambda^1\p_1\right) X^N \\
\nl = \left(2\lambda^0 T\right)^2 \left( \p_1^2X^K +
\Gamma^{K}_{MN}\p_1X^M\p_1X^N \right).\ea

Now we are going to show that by introducing an appropriate ansatz, one
can reduce the classical string dynamics to the dynamics of a massless
particle constrained on a certain surface whenever there exists at least
one Killing vector for the background metric. Indeed, let us split the
index $M=(\mu,a)$, $\{\mu\}\neq\{\emptyset\}$ and let us suppose that
there exist a number of independent Killing vectors $\eta_{\mu}$. Then in
appropriate coordinates $\eta_{\mu}=\frac{\p}{\p x^{\mu}}$ and the metric
does not depend on $X^{\mu}$. In other words, from now on we will work
with the metric \ba\nl g_{MN}=g_{MN}\left(X^a\right).\ea On the other
hand, we observe that \ba\nl X^M(\tau,\sigma) =
F^{M}_{\pm}[w_{\pm}(\tau,\sigma)],\h
w_{\pm}(\tau,\sigma)=(\lambda^1\pm2\lambda^0 T)\tau + \sigma \ea are
solutions of the equations of motion (\ref{ele}) in arbitrary $D$-
dimensional background $g_{MN}\left(X^K\right)$, depending on $D$
arbitrary functions $F^{M}_{+}$ or $F^{M}_{-}$ (see also \cite{MR92}).
Taking all this into account, we propose the ansatz \ba\label{ta}
X^{\mu}\left(\tau,\sigma\right) &=& C^{\mu}_{\pm}w_{\pm} +
y^{\mu}\left(\tau\right),\h C^{\mu}_{\pm}=constants,
\\ \nl  X^{a}\left(\tau,\sigma\right) &=& y^{a}\left(\tau\right).\ea
Inserting (\ref{ta}) into constraints (\ref{c1}) and (\ref{c2}) one
obtains (the overdot is used for $d/d\tau$) \ba\nl g_{MN}(y^a)\dot{y}^M
\dot{y}^N \pm 2\lambda^0 TC^{\mu}_{\pm} \left[g_{\mu N}(y^a)\dot{y}^N \pm
2\lambda^0 TC^{\nu}_{\pm} g_{\mu\nu}(y^a)\right] = 0,\\ \nl C^{\mu}_{\pm}
\left[g_{\mu N}(y^a)\dot{y}^N \pm 2\lambda^0 TC^{\nu}_{\pm}
g_{\mu\nu}(y^a)\right] = 0.\ea Obviously, this system of two constraints
is equivalent to the following one \ba\label{tc1'} g_{MN}(y^a)\dot{y}^M
\dot{y}^N = 0,\\ \label{tc2} C^{\mu}_{\pm} \left[g_{\mu N}(y^a)\dot{y}^N
\pm 2\lambda^0 TC^{\nu}_{\pm} g_{\mu\nu}(y^a)\right] = 0.\ea Using the
ansatz (\ref{ta}) and constraint (\ref{tc2}) one can reduce the initial
Lagrangian to get \ba\nl L^{red}(\tau)\propto \frac{1}{4\lambda^0}
\left[g_{MN}(y^a)\dot{y}^M\dot{y}^N - 2\left(2\lambda^0 T\right)^2
C^{\mu}_{\pm}C^{\nu}_{\pm}g_{\mu\nu}(y^a)\right].\ea It is easy to check
that the constraint (\ref{tc1'}) can be rewritten as $g^{MN}p_M p_N=0$,
where $p_M=\p L^{red}/\p\dot{y}^M$ is the momentum conjugated to $y^M$.
All this means that we have obtained an effective dynamical system
describing a massless point particle moving in a gravity background
$g_{MN}(y^a)$ and in a potential \ba\nl U\propto T^2
C^{\mu}_{\pm}C^{\nu}_{\pm}g_{\mu\nu}(y^a)\ea on the constraint surface
(\ref{tc2}).

Analogous results can be received if one uses the ansatz
\ba\label{sa} X^{\mu}\left(\tau,\sigma\right) =
C^{\mu}_{\pm}w_{\pm} + z^{\mu}\left(\sigma\right),\h
X^{a}\left(\tau,\sigma\right) = z^{a}\left(\sigma\right).\ea
Now putting (\ref{sa}) in (\ref{c1}) and (\ref{c2}) one gets
($'$ is used for $d/d\sigma$)
\ba\nl &&\left[\left(2\lambda^0 T\right)^2 + (\lambda^1)^2\right]
g_{MN}z'^Mz'^N \\ \nl
&&+ 4\lambda^0 T\left[\left(2\lambda^0 T\mp\lambda^1\right)
C^{\mu}_{\pm}g_{\mu N}z'^N
+ 2\lambda^0 TC^{\mu}_{\pm}C^{\nu}_{\pm}
g_{\mu\nu}\right]=0,\\ \nl
&&\lambda^1 g_{MN}z'^Mz'^N + \left[\left(\lambda^1\mp2\lambda^0 T\right)
C^{\mu}_{\pm}g_{\mu N}z'^N \mp 2\lambda^0 TC^{\mu}_{\pm}C^{\nu}_{\pm}
g_{\mu\nu}\right]=0.\ea
These constraints are equivalent to the following ones
\ba\nl  g_{MN}(z^a)z'^Mz'^N =0,\\
\label{sc2'} C^{\mu}_{\pm}\left[g_{\mu N}(z^a)z'^N +
\frac{2\lambda^0 T}{2\lambda^0 T\mp\lambda^1}
C^{\nu}_{\pm}g_{\mu\nu}(z^a)\right]=0.\ea
The corresponding reduced Lagrangian obtained with the help of (\ref{sa}) and
(\ref{sc2'}) is
\ba\nl L^{red}(\sigma)\propto \frac{\left(\lambda^1\right)^2 -
\left(2\lambda^0 T\right)^2}{4\lambda^0}\left[g_{MN}(z^a)z'^Mz'^N -
2\left(\frac{2\lambda^0 T}{2\lambda^0 T\mp\lambda^1}\right)^2
C^{\mu}_{\pm}C^{\nu}_{\pm}g_{\mu\nu}(z^a)\right]\ea
and a similar interpretation can be given as before.

In both cases - the ansatz (\ref{ta}) and the ansatz (\ref{sa}),
the reduced Lagrangians do not depend on $y^\mu$ and $z^\mu$ respectively,
and their conjugated generalized momenta are conserved.

Let us point out that the main difference between tensile and null strings,
from the point of view of the reduced Lagrangians, is the absence of a
potential term for the latter.

Because the consequences of (\ref{ta}) and (\ref{sa}) are similar,
our further considerations will be based on the ansatz (\ref{ta}).

\section{Exact solutions in curved background}
\hspace{1cm} To obtain the equations which we are going to consider, we
use the ansatz (\ref{ta}) and rewrite (\ref{ele}) in the form
\ba\label{le} g_{KL}\ddot{y}^L + \Gamma_{K,MN}\dot{y}^M\dot{y}^N \pm
4\lambda^0 TC^{\mu}_{\pm}\Gamma_{K,\mu N}\dot{y}^N = 0.\ea At first, we
set $K=\mu$ in the above equality. It turns out that in this case the
equations (\ref{le}) reduce to \ba\nl
\frac{d}{d\tau}\left[g_{\mu\nu}\dot{y}^{\nu} + g_{\mu a}\dot{y}^a \pm
2\lambda^0 TC^{\nu}_{\pm}g_{\mu\nu}\right]=0,\ea i.e we have obtained the
following first integrals (constants of the motion) \ba\label{fi}
g_{\mu\nu}\dot{y}^{\nu} + g_{\mu a}\dot{y}^a \pm 2\lambda^0
TC^{\nu}_{\pm}g_{\mu\nu} = A^{\pm}_{\mu} = constants.\ea They correspond
to the conserved momenta $p_\mu$. From the constraint (\ref{tc2}) it
follows that the right hand side of (\ref{fi}) must satisfy the condition
\ba\nl A^{\pm}_{\mu} C^{\mu}_{\pm} = 0.\ea

Using (\ref{fi}), the equations (\ref{le}) for $K=a$ and the constraint
(\ref{tc1'}) can be rewritten as \ba\label{ema}
2\frac{d}{d\tau}\left(h_{ab}\dot{y}^b\right) - \left(\p_a
h_{bc}\right)\dot{y}^b\dot{y}^c + \p_a V^{\pm} = 4\p_{[a}
\left(g_{b]\mu}k^{\mu\nu}A^{\pm}_{\nu}\right)\dot{y}^b\ea and
\ba\label{c2a} h_{ab}\dot{y}^a\dot{y}^b + V^{\pm} =0, \ea where \ba\nl
h_{ab}\equiv g_{ab} - g_{a\mu}k^{\mu\nu}g_{\nu b},\h\h V^{\pm}\equiv
A^{\pm}_{\mu}A^{\pm}_{\nu}k^{\mu\nu} + \left(2\lambda^0
T\right)^2C^{\mu}_{\pm}C^{\nu}_{\pm}g_{\mu\nu},\ea and $k^{\mu\nu}$ is by
definition the inverse of $g_{\mu\nu}$:
$k^{\mu\lambda}g_{\lambda\nu}=\delta^{\mu}_{\nu}$. For example, when
$g_{MN}$ does not depend on the coordinate $y^q$ \ba\nl h_{ab}= g_{ab} -
\frac{g_{aq}g_{qb}}{g_{qq}},\ea when $g_{MN}$ does not depend on two of
the coordinates (say $y^q$ and $y^s$) \ba\nl h_{ab}= g_{ab} -
\frac{g_{aq}g_{ss}g_{qb} - 2g_{aq}g_{qs}g_{sb} +
g_{as}g_{qq}g_{sb}}{g_{qq}g_{ss} - g_{qs}^{2}},\ea and so on.

At this stage, we restrict the metric $h_{ab}$ to be a diagonal one, i.e.
\ba\label{mc} g_{ab} = g_{a\mu}k^{\mu\nu}g_{\nu b}, \h\h \mbox{for}\h\h
a\ne b.\ea This allows us to transform further equations (\ref{ema}) and
obtain (there is no summation over $a$) \ba\label{emat}
\frac{d}{d\tau}\left(h_{aa}\dot{y}^a\right)^2 + \dot{y}^a\p_a\left(h_{aa}
V^{\pm}\right) + \dot{y}^a\sum_{b\ne a}
\left[\p_a\left(\frac{h_{aa}}{h_{bb}}\right)\left(h_{bb}\dot{y}^b\right)^2
- 4\p_{[a}A^{\pm}_{b]} h_{aa}\dot{y}^b\right] = 0,\ea where we have
introduced the notation \ba\label{gp} A^{\pm}_{a} \equiv
g_{a\mu}k^{\mu\nu}A^{\pm}_{\nu}.\ea In receiving (\ref{emat}), the
constraint (\ref{c2a}) is also used after taking into account the
restriction (\ref{mc}).

To reduce the order of the differential equations (\ref{emat}) by one, we
first split the index $a$ in such a way that $y^r$ is one of the
coordinates $y^a$, and $y^{\alpha}$ are the others. Then we impose the
conditions \ba\label{cond}
\p_{\alpha}\left(\frac{h_{\alpha\alpha}}{h_{aa}}\right)=0,\h
\p_{\alpha}\left(h_{rr}\dot{y}^r\right)^2 = 0,\h
\p_{r}\left(h_{\alpha\alpha}\dot{y}^{\alpha}\right)^2 = 0,\h
A^{\pm}_{\alpha}=\p_{\alpha}f^{\pm}.\ea The result of integrations,
compatible with (\ref{c2a}) and (\ref{mc}), is the following \ba\nl
\left(h_{\alpha\alpha}\dot{y}^{\alpha}\right)^2 = D_{\alpha} \left(y^a\ne
y^{\alpha}\right) + h_{\alpha\alpha}\left[2\left( A^{\pm}_{r}-\p_r
f^{\pm}\right)\dot{y}^r - V^{\pm}\right]=E_{\alpha}
\left(y^{\beta}\right),\\ \label{fia} \left(h_{rr}\dot{z}^{r}\right)^2 =
h_{rr}\left\{\left(\sum_{\alpha}-1\right) V^{\pm} -
\sum_{\alpha}\frac{D_{\alpha}}{h_{\alpha\alpha}}\right\} +
\left[\sum_{\alpha}\left(A^{\pm}_{r}-\p_r f^{\pm}\right)\right]^2 =
E_r\left(y^r\right),\ea where $D_{\alpha}$, $E_{\alpha}$, $E_r$ are
arbitrary functions of their arguments, and \ba\nl \dot{z}^r \equiv
\dot{y}^r + \frac{\sum_{\alpha}}{h_{rr}} \left(A^{\pm}_{r}-\p_r
f^{\pm}\right).\ea To find solutions of the above equations without
choosing particular metric, we have to fix all coordinates $y^a$ except
one. If we denote it by $y^A$, then the $exact$ solutions of the equations
of motion and constraints for a string in the considered curved background
are given by \ba\label{esec} X^{\mu}\left(X^A,\s\right)=X^{\mu}_{0} +
C^{\mu}_{\pm}\left(\lambda^1\tau+\sigma\right)-
\int_{X_0^A}^{X^A}k_0^{\mu\nu}\left[g^0_{\nu A}\mp A^{\pm}_{\nu}
\left(-\frac{h^0_{AA}}{V^{\pm 0}}\right)^{1/2}\right]d u ,\\ \nl
X^a=X^a_0=constants\h\mbox{for}\h a\ne A,\h \tau\left(X^A\right)=\tau_0
\pm \int_{X_0^A}^{X^A} \left(-\frac{h^0_{AA}}{V^{\pm 0}}\right)^{1/2}d
u,\ea where $X^{\mu}_{0}$, ${X_0^A}$ and $\tau_0$ are arbitrary constants.
In these expressions \ba\nl h^0_{AA}=h^0_{AA}\left(X^A\right)
=h_{AA}\left(X^A,X^{a\ne A}_0\right)\ea and analogously for $V^{\pm 0}$,
$k_0^{\mu\nu}$ and $g^0_{\nu A}$.

\section{Turning on the B field}
\hspace{1cm} Here we are going to obtain exact string solutions when the
background also includes the Kalb-Ramond antisymmetric gauge field
$B_{MN}(X)$. To this end, we start with the bosonic part of the
Green-Schwarz superstring action
\ba\label{asB} S_1=-\frac{T}{2}\int d^{2}\xi \left[
\sqrt{-\gamma}\gamma^{mn} \p_m X^M\p_n X^N g_{MN}(X) -
\varepsilon^{mn}\p_m X^M\p_n X^N B_{MN}(X)\right].\ea Varying (\ref{asB})
with respect to $X^M$ and $\gamma_{mn}$, we obtain the equations of motion
\ba\nl &&-g_{LK}\left[\p_m\left( \sqrt{-\gamma}\gamma^{mn} \p_n X^K\right)
+ \sqrt{-\gamma}\gamma^{mn}\Gamma^{K}_{MN}\p_m X^M\p_n X^N \right] \\ \nl
&&= \frac{1}{2}H_{LMN}\varepsilon^{mn}\p_m X^M\p_n X^N,\h H=dB, \ea and
the constraints \ba\nl
\left(\gamma^{kl}\gamma^{mn}-2\gamma^{km}\gamma^{ln}\right)\p_m X^M\p_n
X^N g_{MN}\left(X\right)=0.\ea In the gauge $\gamma^{mn} = constants$ and
using (\ref{par}), the Euler-Lagrange equations can be rewritten as
\ba\label{emB} &&g_{LM}\left[\left(\p_0-\lambda^1\p_1\right)^2 -
(2\lambda^0 T)^2\p^2_1\right]X^M \\ \nl &&+
\Gamma_{L,MN}\left[\left(\p_0-\lambda^1\p_1\right)X^M
\left(\p_0-\lambda^1\p_1\right)X^N - (2\lambda^0 T)^2\p_1 X^M \p_1 X^N
\right]\\ \nl &&= 2\lambda^0 T H_{LMN}\p_0 X^M \p_1 X^N ,\ea and the {\it
independent} constraints take the same form (\ref{c1}), (\ref{c2}) as
before. Putting the ansatz (\ref{ta}) into (\ref{emB}) one obtains
\ba\label{remB} g_{KL}\ddot{y}^L + \Gamma_{K,MN}\dot{y}^M\dot{y}^N +
2\lambda^0 TC^{\mu}_{\pm}\left(H_{K\mu N} \pm 2\Gamma_{K,\mu
N}\right)\dot{y}^N = 0.\ea Now we suppose that the tensor field $B_{MN}$
has the same symmetry as the background metric $g_{MN}$, i.e.
$\p_{\mu}g_{MN} = \p_{\mu}B_{MN} = 0$. Then it follows that \ba\label{cmB}
g_{\mu N}\dot{y}^N + 2\lambda^0 TC_{\pm}^{\nu}\left(B_{\mu\nu} \pm
g_{\mu\nu}\right) = B^{\pm}_{\mu} = constants\ea are first integrals of
the equations of motion (\ref{remB}). These conserved quantities are
compatible with the constraint (\ref{tc2}) when $B^{\pm}_{\mu}
C^{\mu}_{\pm} = 0$. Using (\ref{cmB}), the equations of motion for $y^a$
and the other constraint (\ref{tc1'}) can be transformed into
\ba\label{emaB} &&2\frac{d}{d\tau}\left(h_{ab}\dot{y}^b\right) -
\left(\p_a h_{bc}\right)\dot{y}^b\dot{y}^c + \p_a V^{\pm}_B = 4\p_{[a}
B^{\pm}_{c]} \dot{y}^c,\\ \label{c2aB} &&h_{ab}\dot{y}^a\dot{y}^b +
V^{\pm}_B =0, \ea where now \ba\nl &&V^{\pm}_B\equiv \left(2\lambda^0
T\right)^2 C^{\mu}_{\pm}C^{\nu}_{\pm}g_{\mu\nu} + \left(B^{\pm}_{\mu} -
2\lambda^0 TB_{\mu\lambda}C^{\lambda}_{\pm}\right) k^{\mu\nu}
\left(B^{\pm}_{\nu} - 2\lambda^0 TB_{\nu\rho}C^{\rho}_{\pm}\right),\\ \nl
&&B^{\pm}_a \equiv g_{a\mu}k^{\mu\nu}B^{\pm}_{\nu} + 2\lambda^0
T\left(B_{a\lambda}-g_{a\mu}k^{\mu\nu}B_{\nu\lambda}\right)
C^{\lambda}_{\pm}.\ea The comparison of (\ref{emaB}), (\ref{c2aB}) with
(\ref{ema}), (\ref{c2a}) shows that the former can be obtained from the
latter by the replacements \ba\nl V^{\pm}\mapsto V^{\pm}_B,\h A^{\pm}_a
\mapsto B^{\pm}_a ,\ea i.e these equalities are form-invariant. Therefore,
the first integrals (\ref{fia}) will have the same form as before under
the same conditions on the metric $h_{ab}$ and with $B^{\pm}_{\alpha}
=\p_{\alpha}f^{\pm}_{B}$. The corresponding generalization of the exact
solutions (\ref{esec}) for $B_{MN}(X)\ne 0$ will be \ba\label{esecB}
&&X^{\mu}\left(X^A,\s\right)=X^{\mu}_{0} +
C^{\mu}_{\pm}\left(\lambda^1\tau+\sigma\right)\\ \nl &&-
\int_{X_0^A}^{X^A}k_0^{\mu\nu}\left[g^0_{\nu A}\mp\left(B^{\pm}_{\nu} -
2\lambda^0 T B^0_{\nu\lambda}C^{\lambda}_{\pm} \right)
\left(-\frac{h^0_{AA}}{V^{\pm 0}_B}\right)^{1/2}\right]d u ,\\ \nl
&&X^a=X^a_0=constants\h\mbox{for}\h a\ne A,\h \tau\left(X^A\right)=\tau_0
\pm \int_{X_0^A}^{X^A} \left(-\frac{h^0_{AA}}{V^{\pm 0}_B}\right)^{1/2}d
u.\ea

We note that till now we have used the string frame ($\sigma$ - model)
metric. It would be useful to have the obtained result written in Einstein
frame metric, i.e. the frame in which the $D$-dimensional Einstein-Hilbert
action is free from dilatonic scalar factor. In particular, we will need
it in one of the following sections. In $D$ dimensions, the connection
between these two type of metrics is \cite{DKL95} \ba\label{efm}
g_{MN}\equiv g_{MN}^{string}=\exp\left( \frac{a(D)}{2}\phi\right)
g_{MN}^E,\ea where for the string \ba\nl a(D) = \sqrt{\frac{8}{D-2}},\ea
and $\phi(X)$ is the dilaton field. With the help of (\ref{efm}), one can
check that (\ref{esecB}) will give the right formulas for the exact
solution in the Einstein frame, if one takes all metric coefficients in
this frame and replaces $V^{\pm 0}_B$ with \ba\nl\exp\left( \frac{a(D)}{2}
\phi^0\right)V^{\pm 0}_B\ea in the expression for $X^\mu$, and with \ba\nl
\exp\left(-\frac{a(D)}{2} \phi^0\right)V^{\pm 0}_B\ea in the expression
for $\tau$, where $\phi^0 = \phi^0\left(X^A\right)$, and \ba\nl
V^{\pm}_B\equiv &&\left(B^{\pm}_{\mu} - 2\lambda^0
TB_{\mu\lambda}C^{\lambda}_{\pm}\right) k_E^{\mu\nu} \left(B^{\pm}_{\nu} -
2\lambda^0
TB_{\nu\rho}C^{\rho}_{\pm}\right)\exp\left(-\frac{a}{2}\phi\right) +\\ \nl
&&\left(2\lambda^0 T\right)^2
C^{\mu}_{\pm}C^{\nu}_{\pm}g^E_{\mu\nu}\exp\left(\frac{a}{2}\phi\right)
.\ea

Let us finally give the induced metric $G_{mn}$ which arises on the string
worldsheet in our parameterization of the auxiliary metric $\gamma^{mn}$
given by (\ref{par}) and after taking into account the ansatz (\ref{ta}).
It is \ba\nl  G_{00}=\left[\left(\lambda^1\right)^2 - \left(2\lambda^0 T
\right)^2\right]G_{11},\h G_{01}=\lambda^1 G_{11},\h G_{11}=
C^{\mu}_{\pm}C^{\nu}_{\pm}g_{\mu\nu}\left(X^a\right).\ea

\section{Exact D-string solutions}
\hspace{1cm} In this section our aim is to consider the $D$-string
dynamics in nontrivial backgrounds. We will use the action \ba\label{azha}
S_D=-\frac{T_D}{2}\int d^{2}\xi\exp\left(-a\phi\right)
\sqrt{-\mathcal{K}}\mathcal{K}^{mn}\left(G_{mn} + B_{mn} + \left(g_s T_D
\right)^{-1} F_{mn}\right),\ea introduced in \cite{AZH97}, which is
classically equivalent to the Dirac-Born-Infeld action \ba\nl
S_{DBI}=-T_D\int d^{2}\xi \exp\left(-a\phi\right)\sqrt{-\det\left(G_{mn} +
B_{mn} + \left(g_s T_D \right)^{-1} F_{mn}\right)}.\ea The notations used
in (\ref{azha}) are as follows. $T_D=T/g_s$ is the $D$-string tension,
where $T=\left(2\pi\alpha'\right)^{-1}$ is the (fundamental) string
tension and $g_s = \exp\langle\phi\rangle$ is the string coupling
expressed by the dilaton vacuum expectation value $\langle\phi\rangle$.
$G_{mn}(X)= \p_m X^M\p_n X^N g_{MN}(X)$, $B_{mn}(X)= \p_m X^M\p_n X^N
B_{MN}(X)$ and $\phi(X)$ are the pullbacks of the background metric,
antisymmetric tensor and dilaton to the $D$-string worldsheet, while
$F_{mn}(\xi)$ is the field strength of the worldsheet $U(1)$ gauge field
$A_m(\xi)$. $\mathcal{K}$ is the determinant of the matrix
$\mathcal{K}_{mn}$, $\mathcal{K}^{mn}$ is its inverse, and these matrices
have symmetric as well as antisymmetric part \ba\nl
\mathcal{K}^{mn}=\mathcal{K}^{(mn)}+\mathcal{K}^{[mn]}\equiv
\gamma^{mn}_{D} + \omega^{mn},\ea where the symmetric part
$\gamma^{mn}_{D}$ is the analogue of the auxiliary metric $\gamma^{mn}$ in
the string actions (\ref{as}) and (\ref{asB}).

To proceed further, we set in (\ref{azha})\ba\label{parD} \gamma^{mn}_D=
\left(\begin{array}{cc}-1&\lambda^1\\ \lambda^1&-\left(\lambda^1 +
\lambda^2\right)\left(\lambda^1 - \lambda^2\right) + \left(2\lambda^0
T_D\right)^2\end{array}\right),\h \omega^{mn} = -\lambda^2
\varepsilon^{mn},\ea  and obtain the Lagrangian density \ba\nl {\cal
L}_D&=&
\frac{\exp(-a\phi)}{4\lambda^0}\biggl\{g_{MN}\left(\p_0-\lambda^1\p_1\right)
X^M\left(\p_0-\lambda^1\p_1\right) X^N\\ \nl &-& \left[\left( \lambda^2
\right)^2 +\left(2\lambda^0 T_D\right)^2\right]g_{MN}\p_1X^M\p_1X^N \\ \nl
&+& 2\lambda^2\left[B_{MN}\p_0 X^M\p_1 X^N +  \left(g_s T_D \right)^{-1}
F_{01}\right]\biggr\}.\ea The constraints that follow from here are \ba\nl
&&g_{MN} \left(\p_0-\lambda^1\p_1\right) X^M
\left(\p_0-\lambda^1\p_1\right) X^N + \left[\left(2\lambda^0 T_D\right)^2
+ \left( \lambda^2\right)^2\right] g_{MN}\p_1X^M\p_1X^N = 0,\\ \nl
&&g_{MN}\left(\p_0-\lambda^1\p_1\right) X^M\p_1X^N=0,\\ \nl &&B_{MN}\p_0
X^M\p_1 X^N +  \left(g_s T_D \right)^{-1} F_{01}-\lambda^2
g_{MN}\p_1X^M\p_1X^N = 0.\ea The last constraint we will use to express
the worldsheet field strength $F_{01}$ through the background fields
$g_{MN}$,$B_{MN}$ and partial derivatives of the received solutions for
$X^M$. The other constraints we will solve together with the equations of
motion.

As until now, we will work in the gauge in which the Lagrange multipliers
are fixed: $\lambda^{0,1,2}=constants$. In this case, the equations of
motion for $A_m$ \ba\nl \p_m \left[\frac{\lambda^2
\exp\left(-a\phi\right)}{2\lambda^0 g_s T_D}\right] = 0\ea are identically
satisfied when the dilaton is also fixed:$\phi=\phi_0=constant$. The
remaining equations are those for $X^M$ and they are \ba\label{emD}
&&g_{LM}\left[\left(\p_0-\lambda^1\p_1\right)^2-\left[(2\lambda^0 T_D)^2 +
\left( \lambda^2\right)^2\right] \p^2_1\right]X^M \\ \nl &&+
\Gamma_{L,MN}\left[\left(\p_0-\lambda^1\p_1\right)X^M
\left(\p_0-\lambda^1\p_1\right)X^N - \left[(2\lambda^0 T_D)^2 +  \left(
\lambda^2\right)^2 \right]\p_1 X^M \p_1 X^N \right]\\ \nl &&= \lambda^2
H_{LMN}\p_0 X^M \p_1 X^N .\ea

The following step is to apply our ansatz (\ref{ta}). However, in the
$D$-string case it does not work properly - a little modification is
needed. Actually, now the background independent solutions of the
equations of motion (\ref{emD}) are \ba\nl X^M(\tau,\sigma) =
F^{M}_{\pm}[(\lambda^1+A_{\pm})\tau + \sigma],\h
A_{\pm}=\pm\sqrt{(2\lambda^0 T_D)^2 + \left( \lambda^2\right)^2},\ea where
again $F^{M}_{\pm}$ are arbitrary functions of their arguments. Therefore,
the appropriate ansatz is \ba\label{taD} X^{\mu}\left(\tau,\sigma\right) =
C^{\mu}_{\pm}[(\lambda^1+A_{\pm})\tau + \sigma] +
y^{\mu}\left(\tau\right),\h X^{a}\left(\tau,\sigma\right) =
y^{a}\left(\tau\right).\ea The insertion of (\ref{taD}) into the equations
(\ref{emD}) reduce them to the following ones \ba\nl g_{KL}\ddot{y}^L +
\Gamma_{K,MN}\dot{y}^M\dot{y}^N + C^{\mu}_{\pm}\left(\lambda^2 H_{K\mu N}
+ 2A_{\pm}\Gamma_{K,\mu N}\right)\dot{y}^N = 0,\ea which possess the first
integrals \ba\nl g_{\mu N}\dot{y}^N +C_{\pm}^{\nu}\left(\lambda^2
B_{\mu\nu} +A_{\pm}g_{\mu\nu}\right) = B^{\pm}_{\mu} = constants,\h
B^{\pm}_{\mu} C^{\mu}_{\pm} = 0.\ea In full analogy with the previously
considered case, the equations of motion for $y^a$ can be transformed into
the form (\ref{emaB}), where now instead of $V^{\pm}_{B}$ and
$B^{\pm}_{a}$ we have \ba\nl &&V^{\pm}_D\equiv A_{\pm}^2
C^{\mu}_{\pm}C^{\nu}_{\pm}g_{\mu\nu} + \left(B^{\pm}_{\mu} - \lambda^2
B_{\mu\lambda}C^{\lambda}_{\pm}\right) k^{\mu\nu} \left(B^{\pm}_{\nu} -
\lambda^2 B_{\nu\rho}C^{\rho}_{\pm}\right),\\ \nl &&B^{D\pm}_a \equiv
g_{a\mu}k^{\mu\nu}B^{\pm}_{\nu} + \lambda^2
\left(B_{a\lambda}-g_{a\mu}k^{\mu\nu}B_{\nu\lambda}\right)
C^{\lambda}_{\pm}.\ea The corresponding exact solution is \ba\label{esecD}
&&X^{\mu}\left(X^A,\s\right)=X^{\mu}_{0} +
C^{\mu}_{\pm}\left(\lambda^1\tau+\sigma\right)\\ \nl &&-
\int_{X_0^A}^{X^A}k_0^{\mu\nu}\left[g^0_{\nu A}\mp\left(B^{\pm}_{\nu} -
\lambda^2 B^0_{\nu\lambda}C^{\lambda}_{\pm} \right)
\left(-\frac{h^0_{AA}}{V^{\pm 0}_D}\right)^{1/2}\right]d u,\\ \nl
&&X^a=X^a_0=constants\h\mbox{for}\h a\ne A,\h \tau\left(X^A\right)=\tau_0
\pm \int_{X_0^A}^{X^A} \left(-\frac{h^0_{AA}}{V^{\pm 0}_D}\right)^{1/2}d
u,\\ \nl &&\frac{F_{01}}{g_s T_D}=\lambda^2 C^{\mu}_{\pm}g^0_{\mu\nu}
C^{\nu}_{\pm} - \left(B^{\pm}_{\rho}-\lambda^2B^0_{\rho\lambda}
C^{\lambda}_{\pm}\right) k_0^{\rho\mu}B^0_{\mu\nu} C^{\nu}_{\pm}\\ \nl
&&\mp \left(B^0_{A\nu} - g^0_{A\rho}k_0^{\rho\mu}B^0_{\mu\nu}\right)
C^{\nu}_{\pm} \left(-\frac{h^0_{AA}}{V^{\pm 0}_D}\right)^{-1/2} ,\h
\phi=\phi_0=constant.\ea

\section{Examples}
\hspace{1cm} Let us first give an explicit example of exact solution for a
string moving in four dimensional cosmological Kasner type background.
Namely, the line element is ($x^0 \equiv t$) \ba\label{km} ds^2 &=&
g_{MN}dx^M dx^N = -(d t)^2 + \sum_{\mu=1}^{3}t^{2q_{\mu}}(dx^{\mu})^2,\\
\label{Kc} &&\sum_{\mu=1}^{3}q_{\mu}=1,\h \sum_{\mu=1}^{3}q^2_{\mu}=1.\ea
For definiteness, we choose $q_{\mu}=\left(2/3,2/3,-1/3\right)$. The
metric (\ref{km}) depends on only one coordinate $t$, which we identify
with $X^a=y^a(\tau)$ according to our ansatz (\ref{ta}). Correspondingly,
the last two terms in (\ref{emat}) vanish and there is no need to impose
the conditions (\ref{cond}). Moreover, the metric (\ref{km}) is a diagonal
one, so we have $h_{aa}=g_{aa}=-1$. Taking this into account, we obtain
the exact solution of the equations of motion and constraints (\ref{esec})
in the considered particular metric expressed as follows \ba\label{Ks}
X^{\mu}\left(t,\s\right)= X^{\mu}_0 +
C^{\mu}_{\pm}\left(\lambda^1\tau+\sigma\right) \pm
A^{\pm}_{\mu}I^{\mu}(t),\h \tau(t)=\tau_0 \pm I^0(t),\\ \nl I^M (t)\equiv
\int_{t_0}^{t}d u u^{-2q_M}\left(V^{\pm}\right)^{-1/2},\h q_M =
(0,2/3,2/3,-1/3).\ea Although we have chosen relatively simple background
metric, the expressions for $I^M$ are too complicated. Because of that, we
shall write down here only the formulas for the two limiting cases $T=0$
and $T\to\infty$ for $t>t_0\ge 0$. The former corresponds to considering
null strings (high energy string limit).

When $T=0$, $I^M$ reads \ba\nl I^M =
\frac{1}{2\left[\left(A_1^{\pm}\right)^2 +
\left(A_2^{\pm}\right)^2\right]^{1/2}}\Biggl[
\frac{t^{2/3-2q_M}}{\left(q_M-1/3\right)A} F\left(1/2,q_M-1/3;q_M+2/3;
-\frac{1}{A^2t^2}\right)\\ \nl + \frac{t_0^{5/3-2q_M}}{q_M-5/6}
F\left(1/2,5/6-q_M;11/6-q_M;-A^2t_0^2\right) +
\frac{\Gamma\left(q_M-1/3\right)\Gamma\left(5/6-q_M\right)}
{\sqrt{\pi}A^{5/3-2q_M}}\Biggr],\ea where \ba\nl A^2\equiv
\frac{\left(A_3^{\pm}\right)^2}{\left(A_1^{\pm}\right)^2 +
\left(A_2^{\pm}\right)^2},\ea $F\left(a,b;c;z\right)$ is the Gauss'
hypergeometric function and $\Gamma(z)$ is the Euler's $\Gamma$-function.

When $T\to\infty$, $I^M$ is given by the equalities \ba\nl I^{0} &=&
\pm\frac{1}{4\lambda^0TC^3_{\pm}}\Biggl[\frac{6}{C}t^{1/3}
F\left(1/2,-1/6;5/6;-\frac{1}{C^2t^2}\right)\\ \nl
&-&\frac{3}{2}t_0^{4/3}F\left(1/2,2/3;5/3; -C^2t_0^2\right) +
\frac{\Gamma\left(-1/6\right)\Gamma\left(2/3\right)}
{\sqrt{\pi}C^{4/3}}\Biggr],\\ \nl I^{1,2} &=&
\pm\frac{1}{4\lambda^0TC^3_{\pm}}\left[\ln\left|
\frac{\left(1+C^2t^2\right)^{1/2}-1}{\left(1+C^2t^2\right)^{1/2}+1}\right|
- \ln\left|\frac{\left(1+C^2t_0^2\right)^{1/2}-1}
{\left(1+C^2t_0^2\right)^{1/2}+1}\right|\right],\\ \nl I^3 &=&
\pm\frac{1}{2\lambda^0TC^3_{\pm}C^2}\left[\left(1+C^2t^2\right)^{1/2} -
\left(1+C^2t_0^2\right)^{1/2}\right],\ea where \ba\nl C^2\equiv
\frac{\left(C^1_{\pm}\right)^2 + \left(C^2_{\pm}\right)^2}
{\left(C^3_{\pm}\right)^2}.\ea

The solutions for $t<t_0$ may be obtained from the above ones by the
exchange $t\leftrightarrow t_0$.

Let us consider the asymptotic behaviour of the solutions. The tensionless
string solution has the following asymptotics as a function of the cosmic
time $t$ \ba\nl \left|\tau(t)-\tau_0\right| &\buildrel{t \to
0}\over\longrightarrow& \frac{3}{5\left[\left(A_1^{\pm}\right)^2 +
\left(A_2^{\pm}\right)^2\right]^{1/2}}t^{5/3}, \\ \nl
X^{1,2}\left(t,\s\right)&\buildrel{t \to 0}\over\longrightarrow& X^{1,2}_0
+ C^{1,2}_{\pm}\sigma
\mp\frac{3A^{\pm}_{1,2}}{\left[\left(A_1^{\pm}\right)^2 +
\left(A_2^{\pm}\right)^2\right]^{1/2}}t^{1/3}, \\ \nl
X^{3}\left(t,\s\right)&\buildrel{t \to 0}\over\longrightarrow& X^{3}_0 +
C^{3}_{\pm}\left\{\sigma
\mp\frac{3\lambda^1}{5\left[\left(A_1^{\pm}\right)^2 +
\left(A_2^{\pm}\right)^2\right]^{1/2}}t^{5/3}\right\};\ea

\ba\nl \left|\tau(t)-\tau_0\right| &\buildrel{t \to
\infty}\over\longrightarrow& \frac{3}{2\left|A_3^{\pm}\right|}t^{2/3}, \\
\nl X^{1,2}\left(t,\s\right)&\buildrel{t \to\infty}\over\longrightarrow&
X^{1,2}_0 + C^{1,2}_{\pm}\left(\sigma
\mp\frac{3\lambda^1}{2A_3^{\pm}}t^{2/3}\right), \\ \nl
X^{3}\left(t,\s\right)&\buildrel{t \to\infty}\over\longrightarrow& X^{3}_0
+ C^{3}_{\pm}\sigma \mp\frac{3}{4}t^{4/3}.\ea The asymptotic behaviour of
the same solution as a function of the worldsheet time parameter $\tau$ is
given by \ba\nl t(\tau)\equiv X^0(\tau)&\buildrel{\tau\to
\tau_0}\over\longrightarrow& \frac{5}{3}\left[\left(A_1^{\pm}\right)^2 +
\left(A_2^{\pm}\right)^2\right]^{3/10} \left|\tau-\tau_0\right|^{3/5},\\
\nl X^{1,2}\left(\tau,\s\right)&\buildrel{\tau
\to\tau_0}\over\longrightarrow& X^{1,2}_0 + C^{1,2}_{\pm}\sigma \mp
A^{\pm}_{1,2}\left\{\frac{3^4 5
\left|\tau-\tau_0\right|}{\left[\left(A_1^{\pm}\right)^2 +
\left(A_2^{\pm}\right)^2\right]^{2}}\right\}^{1/5} , \\ \nl
X^{3}\left(\tau,\s\right)&\buildrel{\tau \to\tau_0}\over\longrightarrow&
X^{3}_0 + C^{3}_{\pm}\left(\sigma
\mp\lambda^1\left|\tau-\tau_0\right|\right);\ea

\ba\nl X^0(\tau)&\buildrel{\tau\to\infty}\over\longrightarrow&
\left(\frac{2}{3}\left|A_3^{\pm}\right|\tau\right)^{3/2},\\ \nl
X^{1,2}\left(\tau,\s\right)&\buildrel{\tau \to\infty}\over\longrightarrow&
X^{1,2}_0 + C^{1,2}_{\pm}(\sigma + \lambda^1 \tau), \\ \nl
X^{3}\left(\tau,\s\right)&\buildrel{\tau \to\infty}\over\longrightarrow&
X^{3}_0 + C^{3}_{\pm}\sigma \mp\frac{\left(A_3^{\pm}\right)^2}{3}\tau^2.
\ea

In the limit $T\to\infty$, the tensile string solution has the following
behaviour for early times \ba\nl \tau(t)-\tau_0 &\buildrel{t \to
0}\over\longrightarrow& -\frac{3}{8\lambda^0 TC^3_{\pm}}t^{4/3}, \\ \nl
X^{1,2}\left(t,\s\right)&\buildrel{t \to 0}\over\longrightarrow& X^{1,2}_0
+ C^{1,2}_{\pm}\left(\sigma -\frac{3\lambda^1}{8\lambda^0
TC^3_{\pm}}t^{4/3}\right)+ \frac{A^{\pm}_{1,2}}{4\lambda^0 TC^3_{\pm}}
\Gamma(0),
\\ \nl X^{3}\left(t,\s\right)&\buildrel{t \to 0}\over\longrightarrow&
X^{3}_0 + C^{3}_{\pm}\sigma -\frac{3\lambda^1}{8\lambda^0 T}t^{4/3}.\ea
For late times, the asymptotic behaviour is:\ba\nl \tau(t)-\tau_0
&\buildrel{t \to \infty}\over\longrightarrow& \frac{3}{2\lambda^0
TC^3_{\pm}C}t^{1/3}, \\ \nl X^{1,2}\left(t,\s\right)&\buildrel{t
\to\infty}\over\longrightarrow& X^{1,2}_0 + C^{1,2}_{\pm}\left(\sigma +
\frac{3\lambda^1}{2\lambda^0 TC^3_{\pm}C}t^{1/3}\right), \\ \nl
X^{3}\left(t,\s\right)&\buildrel{t \to \infty}\over\longrightarrow&
X^{3}_0 + C^{3}_{\pm}\sigma +\frac{A_3^{\pm}}{2\lambda^0 TC^3_{\pm}C}t.\ea
The asymptotic behaviour of this solution as a function of $\tau$ is given
by  \ba\nl X^0(\tau)&\buildrel{\tau\to \tau_0}\over\longrightarrow&
\left[\frac{8\lambda^0 TC^3_{\pm}}{3}
\left(\tau_0-\tau\right)\right]^{3/4},\\ \nl
X^{1,2}\left(\tau,\s\right)&\buildrel{\tau \to\tau_0}\over\longrightarrow&
X^{1,2}_0 + C^{1,2}_{\pm}\left[\sigma + \lambda^1
\left(\tau-\tau_0\right)\right] + \frac{A^{\pm}_{1,2}}{4\lambda^0
TC^3_{\pm}} \Gamma(0), \\ \nl X^{3}\left(\tau,\s\right)&\buildrel{\tau
\to\tau_0}\over\longrightarrow& X^{3}_0 + C^{3}_{\pm}\left[\sigma +
\lambda^1 \left(\tau-\tau_0\right)\right];\ea

\ba\nl X^0(\tau)&\buildrel{\tau\to\infty}\over\longrightarrow&
\left(\frac{2\lambda^0 TC^3_{\pm}C}{3}\tau\right)^{3},\\ \nl
X^{1,2}\left(\tau,\s\right)&\buildrel{\tau \to\infty}\over\longrightarrow&
X^{1,2}_0 + C^{1,2}_{\pm}(\sigma + \lambda^1 \tau), \\ \nl
X^{3}\left(\tau,\s\right)&\buildrel{\tau \to\infty}\over\longrightarrow&
X^{3}_0 + C^{3}_{\pm}\sigma +\frac{A_3^{\pm}\left(2\lambda^0
TC^3_{\pm}C\right)^2}{27}\tau^3. \ea

The proper string size is \ba\nl l_s \propto
\sqrt{\frac{C^{\mu}_{\pm}C^{\nu}_{\pm}g_{\mu\nu}\left(X^a\right)}{1-\left(
\lambda^1/2\lambda^0 T\right)^2}}.\ea In the considered Kasner space-time,
it grows like $t^{-1/3}$ for $t\to 0$, and like $t^{2/3}$ for
$t\to\infty$. Recall that the space volume depends on the cosmic time $t$
linearly.

Our choice of the scale factors $\left(t^{2/3},t^{2/3},t^{-1/3}\right)$
was dictated only by the simplicity of the solution. However, this is a
very special case of a Kasner type metric. Actually, this is one of the
two solutions of the constraints (\ref{Kc}) (up to renaming of the
coordinates $x^\mu$) for which two of the exponents $q_\mu$ are equal. The
other such solution is $q_{\mu}=\left(0,0,1\right)$ and it corresponds to
flat space-time. Now, we will write down the exact null string solution
$(T=0)$ for a gravity background with arbitrary, but different $q_{\mu}$.
It is given by (\ref{Ks}), where \ba\nl &&I^M(t)= constant
-\frac{\sqrt{\pi}}{A_2^{\pm}}\sum_{k=0}^{\infty}
\frac{\left(A_3^{\pm}/A_2^{\pm}\right)^{2k}}{k!\Gamma\left(1/2-k\right)}
\frac{t^{\mathcal{P}}}{\mathcal{P}}\times \\ \nl &&F\left(1/2+k,\frac{
\mathcal{P}}{2(q_2-q_1)}; \frac{2(q_2-q_3)k+3q_2-2q_1+1-2q_M}{2(q_2-q_1)};
-\left(\frac{A_1^{\pm}}{A_2^{\pm}}\right)^2 t^{2(q_2-q_1)}\right),\\ \nl
&& \mathcal{P}\equiv 2(q_2-q_3)k+q_2+1-2q_M ,\h q_M = (0,q_1,q_2,q_3),\h
\mbox{for $q_1 > q_2$,} \ea  and \ba\nl &&I^M(t)= constant
+\frac{\sqrt{\pi}}{A_1^{\pm}}\sum_{k=0}^{\infty}
\frac{\left(A_3^{\pm}/A_1^{\pm}\right)^{2k}}{k!\Gamma\left(1/2-k\right)}
\frac{t^{\mathcal{Q}}}{\mathcal{Q}}\times \\ \nl &&F\left(1/2+k,\frac{
\mathcal{Q}}{2(q_1-q_2)}; \frac{2(q_1-q_3)k+3q_1-2q_2+1-2q_M}{2(q_1-q_2)};
-\left(\frac{A_2^{\pm}}{A_1^{\pm}}\right)^2 t^{2(q_1-q_2)}\right),\\ \nl
&& \mathcal{Q}\equiv 2(q_1-q_3)k+q_1+1-2q_M,\h\mbox{for $q_1 < q_2$}.\ea
Because there are no restrictions on $q_\mu$, except $q_1\neq q_2\neq
q_3$, the above probe string solution is also valid in generalized Kasner
type backgrounds arising in superstring cosmology \cite{LWC99}. In string
frame, the effective Kasner constraints for the four dimensional
dilaton-moduli-vacuum solution are \ba\nl
&&\sum_{\mu=1}^{3}q_{\mu}=1+\mathcal{K},\h
\sum_{\mu=1}^{3}q^2_{\mu}=1-\mathcal{B}^2,\\ \nl
&&-1-\sqrt{3\left(1-\mathcal{B}^2\right)}\le \mathcal{K}\le
-1+\sqrt{3\left(1-\mathcal{B}^2\right)},\h
\mathcal{B}^2\in\left[0,1\right].\ea In Einstein frame, the metric has the
same form, but in new, rescaled coordinates and with new powers
$\tilde{q}_{\mu}$ of the scale factors. The generalized Kasner constraints
are also modified as follows \ba\nl &&\sum_{\mu=1}^{3}\tilde{q}_{\mu}=1,\h
\sum_{\mu=1}^{3}\tilde{q}^2_{\mu}=1-\tilde{\mathcal{B}}^2 -
\frac{1}{2}\tilde{\mathcal{K}}^2,\h \tilde{\mathcal{B}}^2 +
\frac{1}{2}\tilde{\mathcal{K}}^2\in \left[0,1\right] .\ea Actually, the
obtained tensionless string solution is also relevant to considerations
within a {\it pre-big bang} context, because there exist a class of models
for pre-big bang cosmology, which is a particular case of the given
generalized Kasner backgrounds \cite{LWC99}.

Our next example is for a string moving in the following ten dimensional
supergravity background given in Einstein frame \cite{DKL95} \ba\nl &&ds^2
= g^E_{MN}dx^Mdx^N = \exp(2A)\eta_{mn}dx^mdx^n + \exp(2B)\left(dr^2 +
r^2d\Omega^2_7\right),\\ \nl &&\exp{[-2(\phi-\phi_0)]}= 1+\frac{k}{r^6},\h
\phi_0, k=constants,\\ \nl &&A=\frac{3}{4}(\phi-\phi_0),\h
B=-\frac{1}{4}(\phi-\phi_0),\h B_{01} =
-\exp\left[-2\left(\phi-\frac{3}{4}\phi_0\right)\right]. \ea All other
components of $B_{MN}$ as well as all components of the gravitino $\psi_M$
and dilatino $\lambda$ are zero. If we parameterize the sphere $S^7$ so
that \ba\nl g^E_{10-j,10-j}= \exp(2B)r^2\prod_{l=1}^{j-1}\sin^2x^{10-l},\h
j=2,3,...,7,\h g^E_{99}=\exp(2B)r^2,\ea the metric $g^E_{MN}$ does not
depend on $x^0$, $x^1$ and $x^3$, i.e. $\mu=0,1,3$. Then we set $y^\alpha
= y^\alpha_0 = constants$ for $\alpha=4,...,9$ and obtain a solution of
the equations of motion and constraints as a function of the radial
coordinate $r$: \ba\nl &&X^{\mu}\left(r,\s\right)= X^{\mu}_0 +
C^{\mu}_{\pm}\left(\lambda^1\tau+\sigma\right) \pm I^{\mu}(r),\h
\tau(r)=\tau_0 \pm I(r),\\ \nl &&I^{m}(r)= \eta^{mn} \int_{r_0}^{r}d u
\left[B^{\pm}_{n}+2\lambda^0
T\varepsilon_{nk}C^k_{\pm}\exp\left(-\phi_0/2\right)
\left(1+\frac{k}{u^6}\right)\right]\left(1+\frac{k}{u^6}\right)
W_{\pm}^{-1/2},\\ \nl &&I^{3}(r)= \frac{B_{3}^{\pm}}{s^2}\int_{r_0}^{r}
\frac{d u}{u^2} W_{\pm}^{-1/2},\h s\equiv \prod_{l=1}^{6}\sin y_0^{10-l},
\h I(r)=\exp\left(\phi_0/2\right)\int_{r_0}^{r}d u W_{\pm}^{-1/2},\ea
where \ba\nl &&W_{\pm}=\Biggl\{\left[B^{\pm}_{0}+2\lambda^0 TC^1_{\pm}
\exp\left(-\phi_0/2\right)\left(1+\frac{k}{u^6}\right)\right]^2 \\ \nl &&
- \left[B^{\pm}_{1}-2\lambda^0 TC^0_{\pm}\exp\left(-\phi_0/2\right)
\left(1+\frac{k}{u^6}\right)\right]^2 \Biggr\}
\left(1+\frac{k}{u^6}\right)-\left(\frac{B_3^\pm}{s}\right)^2\frac{1}{u^2}
\\ \nl &&+ \left(2\lambda^0T\right)^2\exp\left(\phi_0\right)
\Biggl\{\left[\left(C^0_\pm\right)^2 - \left(C^1_\pm\right)^2\right]
\left(1+\frac{k}{u^6}\right)^{-1}-\left(C_\pm^3 s\right)^2 u^2
\Biggr\}.\ea This is the solution also in the string frame, because we
have one and the same metric in the action expressed in two different
ways.

The above solution extremely simplifies in the tensionless limit $T\to 0$.
Let us give the manifest expressions for this case. For $r_0<r$, they are:
\ba\nl &&\lim_{T\to 0}I^m (r) = \eta^{mn}B_m^{\pm} \left(J^0 +
kJ^6\right),\h \lim_{T\to 0}I^3 (r) = \frac{B_3^\pm}{s^2}J^2,\h \lim_{T\to
0}I(r) = J^0,\ea where \ba\nl &&J^{\beta}(r) =
-\sqrt{\frac{\pi}{\left(B_0^\pm\right)^2 - \left(B_1^\pm\right)^2}}\\ \nl
&&\times \Biggl\{
\frac{1}{r^{\beta-1}}\sum_{n=0}^{\infty}\frac{\Gamma\left(-
\frac{6n+\beta-5}{4}\right)\left(k/r^6\right)^n}{\left(6n+\beta-1\right)
\Gamma\left( \frac{1-2n}{2}\right)\Gamma\left(
-\frac{2n+\beta-5}{4}\right)} P_{n}^{\left(-\frac{6n+\beta-1}{4},
-n-1\right)} \left(1-2\frac{\delta}{k}r^4\right)
 \\ \nl &&-\frac{1}{r_0^{\beta-1}}\sum_{n=0}^{\infty}\frac{\Gamma\left(
\frac{2n+\beta+3}{4}\right)\left(-\delta/r_0^2
\right)^n}{\left(2n+\beta-1\right) \Gamma\left(
\frac{1-2n}{2}\right)\Gamma\left( \frac{6n+\beta+3}{4}\right)}
P_n^{\left(\frac{2n+\beta-1}{4},
-n-1\right)}\left(1-2\frac{k}{\delta}r_0^{-4}\right)\Biggr\},\\ \nl
&&\delta\equiv \frac{\left(B_3^{\pm}/s\right)^2}{\left(B_0^\pm\right)^2 -
\left(B_1^\pm\right)^2},\ea and $P_n^{(\alpha,\beta)}(z)$ are the Jacobi
polynomials. To obtain the solution for $r_0>r$, one has to exchange $r$
and $r_0$ in the expression for $J^{\beta}$.

Now let us turn to the case of a $D$-string living in five dimensional
{\it anti de Sitter} space-time. The corresponding metric may be written
as \ba\nl &&g_{00}=-\left(1+\frac{r^2}{R^2}\right),\h g_{11}=
\left(1+\frac{r^2}{R^2}\right)^{-1}, \\ \nl &&g_{22}=
r^2\sin^2x^3\sin^2x^4,\h g_{33}=  r^2\sin^2x^4,\h g_{44}=r^2,\ea where $K=
-1/R^2$ is the constant curvature. Now $g_{MN}$ does not depend on $x^0$,
$x^2$ and $B_{MN}=0$. If we fix the coordinates $x^3$, $x^4$ and use the
generic formula (\ref{esecD}), the exact solution as a function of
$r\equiv x^1$ will be \ba\nl &&X^{0}\left(r,\s\right)= X^{0}_0 +
C^{0}_{\pm}\left(\lambda^1\tau+\sigma\right) \mp B^\pm_0\int_{r_0}^{r}d u
\left(1+\frac{u^2}{R^2}\right)^{-1}\left(g_{00}V_D^{\pm0}\right)^{-1/2},
\\ \nl &&X^{2}\left(r,\s\right)= X^{2}_0 +
C^{2}_{\pm}\left(\lambda^1\tau+\sigma\right) \pm
\frac{B^\pm_2}{c^2}\int_{r_0}^{r}\frac{d u}{u^2}
\left(g_{00}V_D^{\pm0}\right)^{-1/2}, \\ \nl &&\tau(r)= \tau_0
\pm\int_{r_0}^{r}d u\left(g_{00}V_D^{\pm0}\right)^{-1/2},\h c\equiv\sin
x_0^3\sin x_0^4,\\ \nl &&F_{01}(r)=-g_sT_D\lambda^2 \Biggl\{\left[\left(
\frac{C_\pm^0}{R}\right)^2 - \left(cC_\pm^2\right)^2 \right]r^2 +
\left(C_\pm^0\right)^2 \Biggr\},\ea where \ba\nl &&g_{00}V_D^{\pm0} =
\left[\left(B_0^\pm\right)^2 - \left(\frac{B^\pm_2}{c R}\right)^2 +
\left(A_\pm C_\pm^0\right)^2\right] - \left( \frac{B^\pm_2}{c}\right)^2
\frac{1}{u^2}\\ \nl &&+A_\pm^2\left[2\left( \frac{C_\pm^0}{R}\right)^2 -
\left(cC_\pm^2\right)^2 \right]u^2 + \left( \frac{A_\pm}{R}\right)^2
\left[\left( \frac{C_\pm^0}{R}\right)^2 - \left(cC_\pm^2\right)^2
\right]u^4.\ea This solution describes a $D$-string evolving in the
subspace ($x^0,x^1,x^2$).

Alternatively, we could fix the coordinates $r=r_0$,
$x^4=x^4_0\equiv\psi_0$ and obtain a solution as a function of the
coordinate $x^3\equiv\theta$. In this case, the result is the following
\ba\nl &&X^{0}\left(\theta,\s\right)= X^{0}_0 +
C^{0}_{\pm}\left(\lambda^1\tau+\sigma\right) \mp \frac{B^\pm_0
\varrho}{g^0_{00}} \int_{\theta_0}^{\theta}d u
\left(-V_D^{\pm0}\right)^{-1/2},
\\ \nl &&X^{2}\left(\theta,\s\right)= X^{2}_0 +
C^{2}_{\pm}\left(\lambda^1\tau+\sigma\right) \pm
\frac{B^\pm_2}{\varrho}\int_{\theta_0}^{\theta}\frac{d u}{\sin^2u}
\left(-V_D^{\pm0}\right)^{-1/2}, \\ \nl &&\tau(\theta)= \tau_0 \pm \varrho
\int_{\theta_0}^{\theta}d u\left(-V_D^{\pm0}\right)^{-1/2},\h
\varrho\equiv r_0\sin\psi_0,\\ \nl &&F_{01}(\theta)=g_sT_D\lambda^2
\left[\left( C_\pm^0\right)^2 g^0_{00} + \left(\varrho C_\pm^2\right)^2
\sin^2\theta\right] ,\ea where \ba\nl -V_D^{\pm0} =
\left[\left(B_0^\pm\right)^2 g_{11}^0 - \left(A_\pm C_\pm^0\right)^2
g^0_{00} \right] -  \left(A_\pm C_\pm^2 \varrho\right)^2\sin^2u -
\frac{\left(B_2^\pm/\varrho\right)^2}{\sin^2u}.\ea This is a solution for
$D$-string placed in the subspace described by the coordinates
($x^0,x^2,x^3$).

Another possibility is when the coordinates $r$ and $\theta$ are kept
fixed, while the coordinate $\psi$ is allowed to vary. However, it is easy
to show that in this case the dependence of the solution on $\psi$ will be
the same as on $\theta$, with some constants changed.

Finally, we will give an example of exact solution for a $D$-string moving
in a non-diagonal metric. To this end, let us consider the ten dimensional
black hole solution of \cite{HMS96}. In string frame metric, it can be
written as \cite{M98} \ba\nl &&ds^2 = \left(1+\frac{r_0^2
\sinh^2\alpha}{r^2}\right)^{-1/2} \left(1+\frac{r_0^2
\sinh^2\gamma}{r^2}\right)^{-1/2}\\ \nl &&\times \Biggl\{-dt^2 + (dx^9)^2
+ \frac{r_0^2}{r^2}\left(\cosh\chi d t+\sinh\chi dx^9\right)^2 \\ \nl &&+
\left(1+\frac{r_0^2 \sinh^2\alpha}{r^2}\right)\left[(dx^5)^2 + (dx^6)^2 +
(dx^7)^2 +(dx^8)^2 \right]\Biggr\}\\ \nl &&+ \left(1+\frac{r_0^2
\sinh^2\alpha}{r^2}\right)^{1/2} \left(1+\frac{r_0^2 \sinh^2\gamma}{r^2}
\right)^{1/2}\left[\left(1-\frac{r_0^2}{r^2}\right)^{-1} dr^2 +
r^2d\Omega_3^2 \right],\\ \label{bhs}
&&\exp\left[-2(\phi-\phi_\infty)\right] = \left(1+\frac{r_0^2
\sinh^2\alpha}{r^2}\right)^{-1} \left(1+\frac{r_0^2 \sinh^2\gamma}{r^2}
\right).\ea The equalities (\ref{bhs}) define a solution of type IIB
string theory, which low energy action in Einstein frame contains the
terms \ba\label{leea} \int d^{10}x\sqrt{-g}\left[R -
\frac{1}{2}\left(\nabla\phi\right)^2 - \frac{1}{12}\exp(\phi)H'^{2}
\right],\ea where $H'$ is the Ramond-Ramond three-form field strength. The
Neveu-Schwarz 3-form field strength, the selfdual 5-form field strength
and the second scalar are set to zero. Therefore, in the solution
(\ref{esecD}), we also have to set $B_{MN}=0$. Besides, our solution
corresponds to a constant dilaton field $\phi=\phi_0$. If we identify
$\phi_0\equiv\phi_\infty$, this leads to $\alpha=\gamma$ in (\ref{bhs}).
On the other hand, we have not included in the $D$-string action
(\ref{azha}) the Ramond-Ramond 2-form gauge field, so we have to set
$H'=0$. It follows from here that $\alpha=\gamma=0$ \cite{M98}. All this
results in a simplification of the metric (\ref{bhs}), and in this
simplified metric the exact $D$-string solution as a function of the
radial coordinate $r$ reads \ba\nl &&X^{\mu}\left(r,\s\right)= X^{\mu}_0 +
C^{\mu}_{\pm}\left(\lambda^1\tau+\sigma\right) \pm I^{\mu}(r),\h \mu =
0,2,5,6,7,8,9, \\ \nl &&X^{3,4}= X_0^{3,4}=constants,\h \tau(r)=\tau_0 \pm
I(r),\ea where \ba\nl &&I^0 = -\int_{r_0}^{r}d u \left[B_0^\pm g_{99} -
B_9^\pm g_{09} \right]g_{11}\mathcal{W}^{-1/2},\\ \nl &&I^9 =
\int_{r_0}^{r}d u \left[B_0^\pm g_{09} - B_9^\pm g_{00}
\right]g_{11}\mathcal{W}^{-1/2},\\ \nl &&I^l= B_l^\pm
\int_{r_0}^{r}\frac{d u}{g_{ll}^0}\mathcal{W}^{-1/2},\h l=2,5,6,7,8,\h
I=\int_{r_0}^{r}d u \mathcal{W}^{-1/2},\\ \nl &&F_{01} = g_s T_D \lambda^2
C^\mu_\pm C^\nu_\pm g_{\mu\nu}^0 (r),\\ \nl &&\mathcal{W}=
\left[\left(B_9^\pm\right)^2 - \frac{\left(A_\pm C_\pm^0\right)^2}{g_{11}}
\right]  g_{00} + \left[\left(B_0^\pm\right)^2 - \frac{\left(A_\pm
C_\pm^9\right)^2}{g_{11}}\right]g_{99} \\ \nl &&- 2\left(B_0^\pm B_9^\pm +
\frac{A_\pm^2 C_\pm^0C_\pm^9}{g_{11}} \right)g_{09} \\ \nl
&&-\frac{1}{g_{11}}\Biggl\{ \sum_{l=5}^{8} \left[\left(A_\pm
C_\pm^l\right)^2 + \left(B_\pm^l\right)^2\right] + \left(
\frac{B_\pm^2}{c}\right)^2 \frac{1}{u^2} + \left(A_\pm C_\pm^2 c\right)^2
u^2 \Biggr\}. \ea

Actually, after fixing the dilaton to constant and $H'$ to zero, we have
obtained a solution for a $D$-string moving in ten dimensional Einstein
gravity background, as is seen from (\ref{leea}). That is why, now we are
going to consider the case $\phi=\phi_0$, $H'\ne 0$. To this aim, we have
to add a Wess-Zumino term to the action (\ref{azha}), describing the
coupling of the $D$-string to the Ramond-Ramond two-form gauge potential
$B'_{MN}$ ($H'=dB'$). It can be shown that this leads to the following
change in the equations of motion (\ref{emD}) \ba\nl \lambda^2 H_{LMN}
\longrightarrow \lambda^2 H_{LMN} + 2\lambda^0 \mu
\exp\left(a\phi_0\right)H'_{LMN},\ea where $\mu$ is the $D$-string charge
($\mu = \pm T_D$ from the requirements of space-time supersymmetry and
worldsheet $\kappa$-invariance of the super $D$-string action). Then one
proceeds as before to obtain an exact solution of the equations of motion
and constraints. This solution is given by (\ref{esecD}), where the
replacement \ba\nl \lambda^2 B_{MN} \longrightarrow \lambda^2 B_{MN} +
2\lambda^0 \mu \exp\left(a\phi_0\right)B'_{MN},\ea must be done. Now we
can put there the background (\ref{bhs}) with $\alpha=\gamma\ne0$ to
obtain an explicit probe $D$-string solution.

\section{Discussion}
\hspace{1cm}In this article we performed some investigation on the
classical string dynamics in $D$-dimensional (super)gravity background. In
Section 2 we begin with rewriting the string action for curved background
in a form in which the limit $T\to 0$ could be taken to include also the
null string case, known to be a good approximation for string dynamics in
strong gravitational fields. Then we propose an ansatz, which reduces the
initial dynamical system depending on two worldsheet parameters
($\tau,\sigma$) to the one depending only on $\tau$, whenever the
background metric does not depend at least on one coordinate. An
alternative ansatz is also given, which leads to a system depending only
on $\sigma$.

Let us note that the usually used conformal gauge for the auxiliary
worldsheet metric $\gamma^{mn}$ corresponds to $\lambda^1 = 0$,
$2\lambda^0 T = 1$. However, the latter does not allow for unified
description of both tensile and tensionless strings.

The ansatz (\ref{ta}) and the ansatz (\ref{sa}) generalize the ones used
in \cite{dVE93}-\cite{SM96} for finding exact string solutions in curved
backgrounds. It is also worth noting that (\ref{ta}) and (\ref{sa}) are
based on the obtained string solutions $F^M_\pm(w_\pm)$, which do not
depend on the background metric. In conformal gauge, they reduce to the
solutions for left- or right-movers, known to be the only {\it background
independent} non-perturbative solutions for an arbitrary static metric,
which are stable and have a conserved topological charge being therefore
topological solitons \cite{MR92}.

In Section 3, using the existence of an abelian isometry group $G$
generated by the Killing vectors $\p/\p x^{\mu}$, the problem of solving
the equations of motion and two constraints in $D$-dimensional curved
space-time $\mathcal{M}_D$ with metric $g_{MN}$ is reduced to considering
equations of motion and one constraint in the coset $\mathcal{M}_D/G$ with
metric $h_{ab}$. As might be expected, an interaction with an effective
gauge field appears in the Euler-Lagrange equations. In this connection,
let us note that if we write down $A^{\pm}_{a}$, introduced in (\ref{gp}),
as \ba\nl A^{\pm}_{a}=A_a^{\nu}A^{\pm}_{\nu},\ea this establishes a
correspondence with the usual Kaluza-Klein type notations and \ba\nl
g_{MN}dy^M dy^N = h_{ab}dy^a dy^b + g_{\mu\nu} \left(dy^\mu + A^\mu_a
dy^a\right)\left(dy^\nu + A^\nu_b dy^b\right).\ea In the remaining part of
Section 3, we impose a number of conditions on the background metric,
sufficient to obtain exact solutions of the equations of motion and
constraints. These conditions are such that the metric is general enough
to include in itself many interesting cases of curved backgrounds in
different dimensions.

In Section 4, the previous results are generalized to include a nontrivial
Kalb-Ramond background gauge field $B_{MN}$, which arises in the
supergravity theories - the low energy limits of superstring theories. The
$B_{MN}$ is restricted to depend on these coordinates on which the
background metric does. There are also indirect restrictions on $g_{MN}$
and $B_{MN}$. They follow from the condition on the effective one-form
potential $B_a^\pm$ in the equations of motion (\ref{emaB}) to be oriented
along one of the coordinate axes, all other components being pure gauges.
It is explained how the derived probe string solution looks like also in
Einstein frame metric - the one used when searching for brane solutions of
the string-theory effective field equations. At the end of the section,
the induced worldsheet metric is given and it depends only on the
$g_{\mu\nu}$ part of the background metric, which corresponds to its
Killing vectors $\p/\p x^\mu$.

In section 5, we consider the $D$-string dynamics. In an appropriate
parameterization of the auxiliary worldsheet field $\mathcal{K}^{mn}$ and
with the help of a modified ansatz, we succeeded to reduce the task of
finding exact solutions to the application of methods used for this
purpose in the previous section. In that case, the limit $T_D \to 0$ also
could be taken and this will give a solution of the equations of motion
and constraints for a {\it tensionless} $D$-string. However, the
tensionless $D$-string is not a null string, i.e. the induced worldsheet
metric is not degenerate. Really, the induced metric is \ba\nl
&&G_{00}=\left[\left(\lambda^1\right)^2 - \left(\lambda^2\right)^2 -
\left(2\lambda^0 T_D \right)^2\right]G_{11},\\ \nl &&G_{01}=\lambda^1
G_{11},\h G_{11}= C^{\mu}_{\pm}C^{\nu}_{\pm}g_{\mu\nu}\left(X^a\right),\\
\nl &&\det G_{mn} = -\left[\left(\lambda^2\right)^2 + \left(2\lambda^0 T_D
\right)^2\right]G^2_{11} .\ea Consequently, when $T_D = 0$, $\det G_{mn}$
is still different from zero. Moreover, the tensile fundamental string can
be viewed as a particular case of the tensionless $D$-string \cite{LU97}.
Indeed, it is not difficult to show by using (\ref{par}) and (\ref{parD})
that the fundamental string action (\ref{asB}) can be obtained from the
$D$-string action (\ref{azha}) by setting $A_m = 0$, $\phi = 0$,
$\lambda^2 = 2\lambda^0 T$ and taking $T_D \to 0$.

The next section is devoted to four examples of exact string solutions in
different backgrounds in four, five and ten dimensions. As an example of
solution in cosmological type background, we consider four dimensional
Kasner space-time. The next example is for a string moving in ten
dimensional supergravity background. The $D$-string solutions are
illustrated with two examples - in five dimensional anti de Sitter and in
ten dimensional black hole backgrounds. It is evident from the solution
(\ref{esecB}) and from the first two examples that the (exact) {\it null}
string dynamics in curved backgrounds, \cite{RZ95} - \cite{DL96},
\cite{DL97}, \cite{PP97} - \cite{JM992}, is simpler than the tensile
string one. Moreover, the tensionless string does not interact with the
$B_{MN}$ background, according to the action (\ref{asB}). This is a
consequence of the condition for $\kappa$-invariance of the Green-Schwarz
superstring action, which bosonic part is (\ref{asB}). Actually, the null
super $p$-brane actions are $\kappa$-invariant in flat space-time without
Wess-Zumino terms \cite{BZ93,B992} . The $D$-string case is different. The
{\it tensionless} $D$-string does interact with the Kalb-Ramond
background.

Let us finally note some specific features of the received solutions. The
parameterization of the worldsheet fields $\gamma^{mn}$ and
$\mathcal{K}^{mn}$ is such that permit for a unified description of the
tensile and tensionless string solutions. The ansatz (\ref{ta}) and the
ansatz (\ref{taD}) contain the terms $\pm 2\lambda^0 TC_\pm^\mu \tau$ and
$A_\pm C_\pm^\mu \tau$ respectively, which disappear in the expressions
(\ref{esecB}) and (\ref{esecD}) for the exact $(D)$-string solutions as a
function of one of the coordinates. There, they are replaced by the
corresponding integrals. The other part of the background independent
solution, $C^{\mu}_{\pm}\left(\lambda^1\tau+\sigma\right)$, is a
particular case of such solution for the null $p$-branes \cite{B993,B995}.
Another distinguishing feature of our exact solutions is that we do not
restrict ourselves to the usually used {\it static gauge} $X^m(\xi) =
\xi^m$, when probe brane dynamics is investigated. It is possible to
impose this gauge on the solutions and this will result in additional
conditions on them.

\vspace*{1cm}
{\bf Acknowledgment}

This work was supported in part by a Shoumen University grant under
contract $No.20/2001$.

\newpage


\begin{thebibliography}{}

\bibitem{dVS87} H. de Vega and N.S\'anchez, Phys. Lett. B {\bf 197}, 320
(1987). 
\bibitem{GSV91} M. Gasperini, N. S\'anchez, and G. Veneziano,
Int. J. Mod. Phys. A {\bf 6}, 3853 (1991); Nucl. Phys. B {\bf 364}, 365
(1991).
\bibitem{dVN92} H. de Vega and A. Nicolaidis, Phys. Lett. B {\bf 295}, 214
(1992). 
\bibitem{dVGN94}H. de Vega, I. Giannakis, and A. Nicolaidis,
Mod. Phys. Lett A {\bf 10}, 2479 (1995). 
\bibitem{CLS96}C. Lousto and N. S\'anchez, Phys. Rev. D {\bf 54}, 6399
(1996). 
\bibitem{MR92} O. Mattos and V. Rivelles, Phys. Rev. Lett. {\bf 70}, 1583
(1993).
\bibitem{dVS93}H. de Vega and N. S\'anchez, Phys. Rev. D {\bf 47}, 3394 (1993).
\bibitem{CdVMS93} F. Combes, H. J. de Vega, A. V. Mikhailov, and N. S\'anchez,
Phys. Rev. D {\bf 50}, 2754 (1994).
\bibitem{dVS94}H. de Vega and N. S\'anchez, Phys. Rev. D {\bf 50},7202
(1994). 
\bibitem{FHL96}A. Frolov, S. Handy, and A. Larsen, Nucl. Phys. B {\bf 468},
336 (1996). 
\bibitem{MO00}J. Maldacena and H. Ooguri,
{\it Strings in $AdS_3$ and the SL(2,R) WZW Model. Part 1: The spectrum},
hep-th/0001053.
\bibitem{dVE93}H. de Vega and I. Egusquiza, Phys. Rev. D {\bf 49}, 763
(1994). 
\bibitem{L93}A. Larsen, Phys. Rev. D {\bf 50}, 2623 (1994).
\bibitem{dVLS93}H. de Vega, A. Larsen, and N. S\'anchez, Nucl. Phys. B
{\bf 427}, 643 (1994).
\bibitem{L94}A. Larsen, Phys. Rev. D {\bf 51}, 4330 (1995).
\bibitem{LS94}A. Larsen and N. S\'anchez, Phys. Rev. D {\bf 50}, 7493 (1994).
\bibitem{dVLS94}H. de Vega, A. Larsen, and N. S\'anchez, Phys. Rev. D
{\bf 51}, 6917 (1995). 
\bibitem{LS952}A. Larsen and N. S\'anchez, Int. J. Mod. Phys. A {\bf 11},
4005 (1996). 
\bibitem{LS96}A. Larsen and N. S\'anchez, Phys. Rev. D {\bf 54}, 2801 (1996).
\bibitem{DL97}M. Dabrowski and A. Larsen, Phys. Rev. D {\bf 57}, 5108 (1998).
\bibitem{dVLS98}H. de Vega, A. Larsen, and N. S\'anchez, Phys. Rev. D
{\bf 58}, 026001 (1998).
\bibitem{LS98}A. Larsen and N. S\'anchez, Phys. Rev. D {\bf 58}, 126002
(1998).
\bibitem{LN98}A. Larsen and A. Nicolaidis, Phys. Rev. D {\bf 60}, 024012
(1999). 
\bibitem{FL99}A. Frolov and A. Larsen,  Class. Quant. Grav. {\bf 16}, 3717
(1999).
\bibitem{B995}P. Bozhilov, Phys. Rev. D {\bf 62}, 105001 (2000).
\bibitem{LS00}A. Larsen and N. S\'anchez, Phys. Rev. D {\bf 62}, 046003
(2000).
\bibitem{LS951}A. Larsen and N. S\'anchez, Phys. Rev. D {\bf 51}, 6929
(1995). 
\bibitem{K95}S. Kar, Phys. Rev. D {\bf 52}, 2036 (1995).
\bibitem{FS95}A. Frolov and A. Larsen, Nucl. Phys. B {\bf 449}, 149
(1995).
\bibitem{FHL95}A. Frolov, S. Handy, and A. Larsen, Phys. Rev. D {\bf 54},
2483 (1996). 
\bibitem{dVE96}H. de Vega and I. Egusquiza, Phys. Rev. D {\bf 54}, 7513
(1996).
\bibitem{SM96}S. Mahapathra, Phys. Rev. D {\bf 55}, 6403 (1997).
\bibitem{dVMMS95}H. de Vega, J. Mittelbrunn, M. Medrano, and N. S\'anchez,
Phys. Rev. D {\bf 52}, 4609 (1995).
\bibitem{dVE95}H. de Vega and I. Egusquiza, Class. Quant. Grav. {\bf 13},
1041 (1996).
\bibitem{DKL95}M. Duff, R. Khuri, and J. Lu, Phys. Rep. {\bf 259}, 213
(1995).
\bibitem{AZH97}M. Abou Zeid and C. Hull, Phys. Lett. B {\bf 404}, 264 (1997).
\bibitem{LWC99}J. Lidsey, D. Wands and E. Copeland, Phys. Rep. {\bf 337},
343 (2000). 
\bibitem{HMS96}G. Horovitz, J. Maldacena, and A. Strominger, Phys. Lett. B
{\bf 383}, 151 (1996). 
\bibitem{M98}J. Maldacena, {\it Black Holes and $D$-branes},
in {\it 1997 Summer School in High Energy Physics and Cosmology}, (Word
Scientific, Singapore, 1998).
\bibitem{LU97}U. Lindstrom and R. von Unge, Phys. Lett. B {\bf 403}, 233
(1997). 
\bibitem{RZ95} S. Roshchupkin, A. Zheltukhin, Class. Quant. Grav.
{\bf 12}, 2519 (1995).
\bibitem{K96} S. Kar, Phys. Rev. D {\bf 53}, 6842 (1996).
\bibitem{DL96} M. Dabrowski, A. Larsen, Phys. Rev. D {\bf 55}, 6409 (1997).
\bibitem{PP97} P. Porfyriadis, D. Papadopoulos, Phys. Lett. B {\bf 417}, 27
(1998). 
\bibitem{GKKP98} I. Giannakis, K. Kleidis, K. Kuiroukidis, D.
Papadopoulos, Mod. Phys. Lett. A {\bf 13}, 3169 (1998).
\bibitem{KP99} A. Kuiroukidis, D. Papadopoulos, Gen. Rel. Grav. {\bf 32},
593 (2000). 
\bibitem{B994} P. Bozhilov, B. Dimitrov,  Phys. Lett. B {\bf 472}, 54 (2000).
\bibitem{JM992} H. Jassal, A. Mukherjee, Phys. Rev. D {\bf 62}, 067501
(2000).
\bibitem{BZ93} I. Bandos, A. Zheltukhin, Fortsch. Phys. {\bf 41}, 619 (1993).
\bibitem{B992} P. Bozhilov, Mod. Phys. Lett. A {\bf 14}, 1335 (1999).
\bibitem{B993}P. Bozhilov, Phys. Rev. D {\bf 60}, 125011 (1999).

\end{thebibliography}
\end{document}